\documentstyle[epsf,12pt]{article}
\newcommand{{\y}}{\'{\i}}

\begin{document}

\begin{center}
{\bf COMPENSATION TEMPERATURE OF A MIXED ISING FERRIMAGNETIC MODEL IN THE
PRESENCE OF EXTERNAL MAGNETIC FIELDS.}\\
\vskip 15 truept
G.~M.\ BUENDIA$^*$, E.\ MACHADO$^*$, M.~A.\ NOVOTNY$^{**}$
\end{center}

\noindent *Departamento de F{\y}sica. Universidad Sim\'on Bol{\y}var, Apartado
89000, Caracas 1080, Venezuela.

\noindent **Supercomputer Computations Research Institute, Florida State
University, Tallahassee, FL 32306-4130, USA. 

\vskip 15 truept
\noindent {\bf ABSTRACT}

The behavior of the compensation temperature of a mixed Ising ferrimagnetic
system on a square lattice in which the two interpenetrating square sublattices
have spins $\sigma$ ($\pm 1/2$) and spins $S$ ($\pm 1,0$)
has been studied with Monte Carlo
methods. Our model includes nearest and  next-nearest neighbor interactions,
a crystal field and an external magnetic field. This model is relevant for
understanding bimetallic molecular ferrimagnetic materials. We found that there
is a narrow range of parameters of the Hamiltonian for which the model has
compensation temperatures and that the compensation point exists only for small
values of the external field.

\vskip 15 truept
\noindent {\bf INTRODUCTION}

Ferrimagnetic ordering plays a crucial role in stable 
crystaline room-temperature magnets, that are currently being synthesized by
several experimental groups in search for materials with technological
applications [1]. In a ferrimagnet the different temperature dependence of the
sublattices magnetizations raises the possibility of the appearance of
compensation temperatures: temperatures below the critical point, where the
total magnetization is zero [2]. The temperature dependence of the
coercivity at the compensation point has important applications in the
field of thermomagnetic recording [3].

Mixed Ising systems are good models to study ferrimagnetic ordering [4]. 
Recent results show that these models can have compensation points when 
their Hamiltonian includes second neighbor interactions [5]. 
These studies have been performed in zero
magnetic field. In this work we study the effect of a constant
external magnetic field on the behavior of the compensation 
temperature, $T_{\rm comp}$. 

\vskip 15 truept
\noindent {\bf THE MIXED ISING MODEL}

Our model consists of two interpenetrating square sublattices. One sublattice
has spins $\sigma$ that can take two values $\pm 1/2$, the other sublattice
has spins $S$ that can take three values, $\pm 1,0$. Each $S$ spin has only
$\sigma$ spins as nearest neighbors and vice versa.

The Hamiltonian of the model is given by,
\begin{equation}
H=-J_1\sum_{\langle nn\rangle} \sigma_i S_j - 
J_2\sum_{\langle nnn\rangle} \sigma_i \sigma_k
- J_3\sum_{\langle nnn\rangle} S_j S_l + 
D\sum_j S_j^2 - h\left(\sum_i \sigma_i + \sum_j S_j\right)
\label{hamilt}
\end{equation}
where the $J$'s are exchange interaction parameters, $D$ is the crystal
field, and $h$ is the external field, all in energy units. We choose 
$J_1$$=$$-1$
such that the coupling between nearest neighbors is antiferromagnetic.

Previous results with Monte Carlo and Transfer Matrix techniques have shown
that the $J_1$$-$$D$ model ($J_2$, $J_3$ and $h$ are all zero) does not have a
compensation temperature [6]. 
These previous studies showed that a compensation
temperature is induced by the presence of the next-nearest neighbor
(nnn) ferromagnetic interaction,
$J_2$, between the $\pm 1/2$ spins. The minimum strength of the $J_2>0$
interaction for a compensation point to appear depends on the other
parameters of the Hamiltonian [5]. In this work we study the
effect of the external field and the $J_3$ parameter on the compensation
temperature.

\vskip 15 truept
\noindent {\bf MONTE CARLO CALCULATIONS}

We use standard importance sampling techniques [7] to simulate the model
described by Eq.~(\ref{hamilt}) on $L$$\times$$L$ 
square lattices with periodic
boundary conditions and $L$$=$$40$. 
Data were generated with $5$$\times$$10^4$ Monte Carlo steps
per site after discarding the first $5$$\times$$10^3$ steps. 
The error bars were
taken from the standard deviation of blocks of 500 sites. We define
$\beta$$=$$1/k_{\rm B}T$ and take Boltzmann's constant $k_{\rm B}$$=$$1$. 
Our program
calculates the internal energy per site, specific heat, the sublattice
magnetizations per site, $M_1$ and $M_2$ defined as,
\begin{equation}
M_1=\frac{2}{L^2}\langle\sum_j S_j\rangle\ \ ,\ \ 
M_2=\frac{2}{L^2}\langle\sum_i \sigma_i\rangle
\end{equation}
and the total magnetization per spin, $M$$=$$\frac{1}{2}(M_1+M_2)$.
The averages are taken over all the configurations, the sums over $j$ are over
all the sites with $S$ spins and the sums over $i$ are over all the sites with
$\sigma$ spins. Each sum has $L^2/2$ terms.

The compensation point, $T_{\rm comp}$, is defined as the point where the two
sublattice magnetizations cancel each other such that the total 
magnetization is zero, i.e.,                                         
\begin{equation}
|M_1(T_{\rm comp})|=|M_2(T_{\rm comp})|
\label{cond1}
\end{equation}
and 
\begin{equation}
sign[M_1(T_{\rm comp})]=-sign[M_2(T_{\rm comp})]
\label{cond2}
\end{equation}
with $T_{\rm comp}$$<$$T_{\rm c}$.  
Note that at the compensation temperature the sublattice magnetizations are
not zero, whereas at the critical temperature, $T_{\rm c}$, 
the total magnetization is zero
and both sublattice magnetizations are also zero.

\vskip 15 truept
\noindent {\bf RESULTS}

\begin{figure}
\epsfxsize=12cm
\centerline{\epsffile{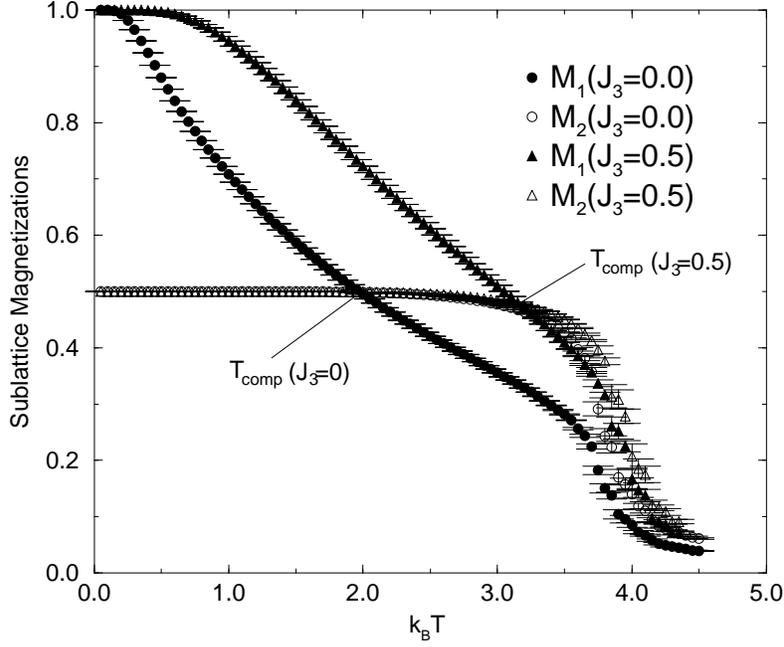}}
\caption{Dependence of the absolute values of the sublattice magnetizations 
with the temperature for $J_3$$=$$0$ (circles) and $J_3$$=$$0.5$ (triangles).  
Here $J_2$$=$$6$, $D$$=$$1$ and $h$$=$$0$.}
\label{fig1}
\end{figure}

Previous studies on the $J_1$$-$$J_2$$-$$D$ 
model showed that, for a fixed value of the
parameters $J_1$ and $D$, there is a minimum value of $J_2$ for which the model
has a compensation point. However, once this minimum value is reached, the
compensation temperature remains almost independent of $J_2$ [5]. In this study
we show that for a fixed value of $J_1$, $D$, and $J_2$, the compensation
temperature can be changed by including the $J_3$ interaction (between the $S$
spins, nnn in the lattice). 
The effect of the ferromagnetic $J_3$ parameter is to increase the
value of the compensation temperature, such that as $J_3$ increases the
compensation temperature approaches the critical temperature and eventually
disappears. In Fig.~\ref{fig1}
we show the absolute values of the sublattice magnetizations for a
$J_1$$-$$J_2$$-$$D$ model ($J_3$$=$$0$, $h$$=$$0$)
and for a $J_1$$-$$J_2$$-$$J_3$$-$$D$ model ($h$$=$$0$). 
Notice that the main effect of the $J_3$
parameter is to keep the $S$ sublattice ordered at higher temperatures, 
such that
the crossing point between both sublattices
[the one that satisfies Eq.~(\ref{cond1}) and Eq.~(\ref{cond2})]
occurs at higher temperatures. As $J_3$ becomes larger, the
compensation temperature
increases toward the critical point.  When both temperatures become equal 
we can not
talk about a compensation point anymore and we only have a critical temperature.

\begin{figure}
\epsfxsize=12cm
\centerline{\epsffile{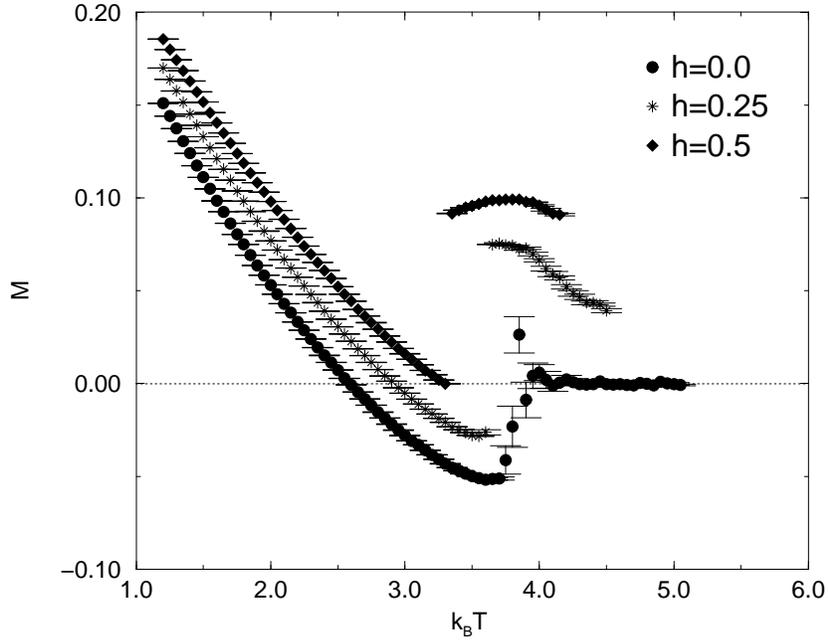}}
\caption{Total magnetizations vs.\ temperature for different values of $h$.  
Here $J_3$$=$$0.25$, $J_2$$=$$6$ and $D$$=$$1$.}
\label{fig2}
\end{figure}

\begin{figure}
\epsfxsize=12cm
\centerline{\epsffile{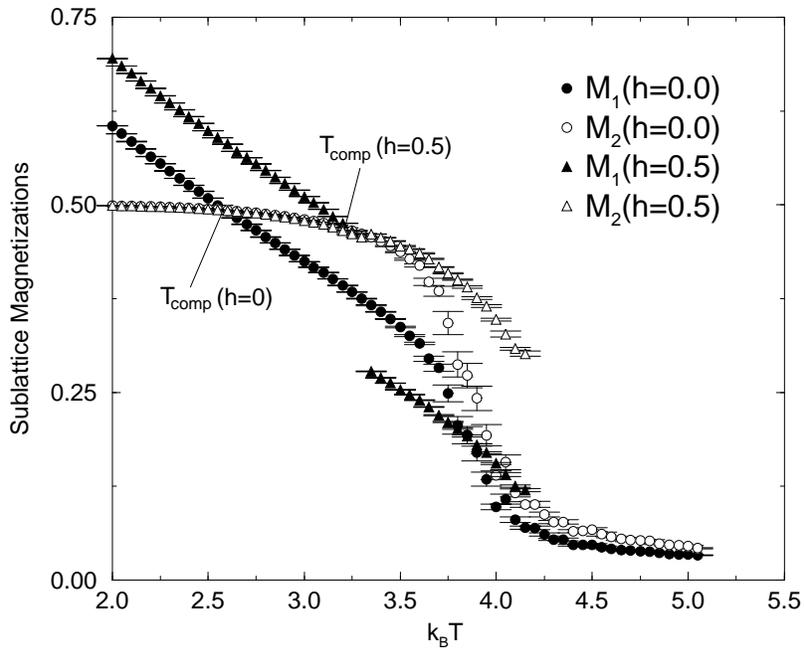}}
\caption{Dependence of the absolute values of the sublattice magnetizations 
with the temperature for $h$$=$$0$ (circles) and $h$$=$$0.5$ (triangles).  
Here $J_3$$=$$0.25$, $J_2$$=$$6$ and $D$$=$$1$.}
\label{fig3}
\end{figure}

When an external field $h$ is added, the compensation temperature 
increases with the field until it disappears, i.e. becomes equal to the 
critical temperature, for
a strong enough value of $h$, as shown in Fig.~\ref{fig2},
where we plot the total magnetization vs. the temperature for several
values of $h$.
The effect of the external field on the compensation temperature is similar
to that due to $J_3$.  Notice that when $h$ is present the system
has a discontinuity in the magnetization that may signal a first order
phase transition. This discontinuity seems to be due almost entirely to a
discontinuity in the magnetization of the $S$ sublattice as shown in
Fig.~\ref{fig3}.

\begin{figure}
\epsfxsize=12cm
\centerline{\epsffile{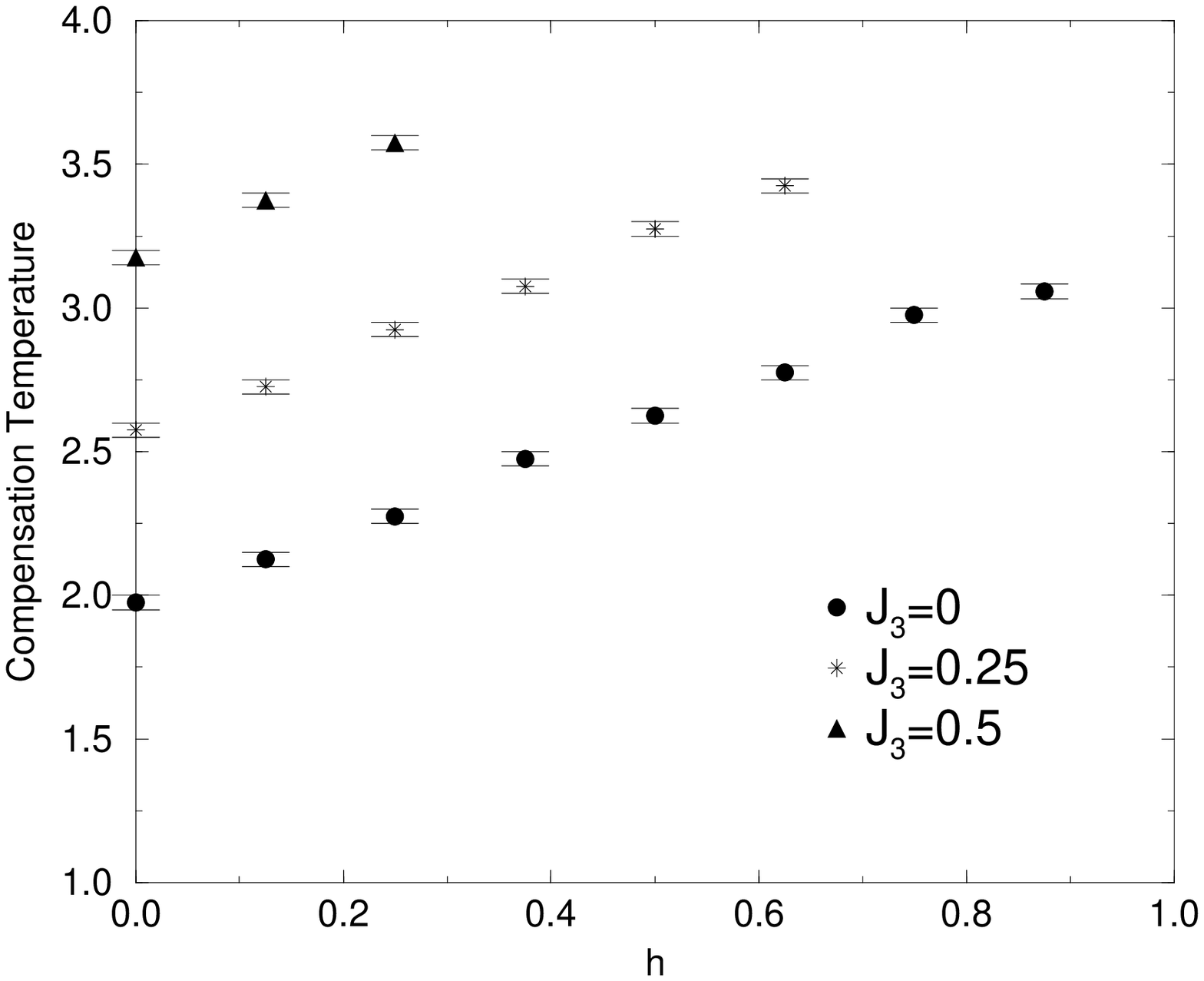}}
\caption{Dependence of the compensation temperature with
the external field for different values of $J_3$. The last point in each curve
was calculated at the highest value of $h$ for which there is a
compensation point for that particular value of $J_3$. 
Here $J_2$$=$$6$ and $D$$=$$1$.}
\label{fig4}
\end{figure}

In Fig.~\ref{fig4}
we show the value of the compensation temperature vs.\ $h$ for different
values of $J_3$. It is interesting to note that for $J_3$ fixed the 
compensation
temperature increases almost linearly with the field until it vanishes. Also
the compensation temperature increases almost 
linearly with $J_3$ for a fixed value of
$h$. As $J_3$ increases the compensation point only exists 
for a very weak or zero external field.

\vskip 15 truept
\noindent {\bf CONCLUSIONS}

There is a strong dependence of the 
compensation temperature on the parameters
in the Hamiltonian, and only a narrow range of parameters for which a
compensation point can exist. The next-nearest neighbor interaction between
the $S$ spins and the
external magnetic field tend to increase the compensation temperature until
it coincides with the critical point, then we no longer have a
compensation point. Since
the compensation temperature is important in several technological 
applications,
such as thermomagnetic recording, it is important to take 
into account that the
external magnetic fields modify the value of the compensation temperature to
higher values and that for strong fields there is no compensation temperature.
Also, the presence of magnetic fields seems to induce a 
discontinuity in the magnetization of the $S$ sublattice.

\vskip 15 truept
\noindent {\bf CONCLUSIONS}

This research was supported in part by U.S.\ NSF Grant No.\ 9520325,
and by the Florida State University 
Supercomputer Computations Research Institute 
(DOE Contract No.\ DE-FC05-85ER25000).
Supercomputer access provided by the DOE at NERSC.

\vskip 15 truept
\noindent {\bf REFERENCES}
\vskip 5 truept
\noindent {\bf 1.} T.~Mallah, S.~Tiebaut, M.~Verdager and P.~Veillet, Science
{\bf 262}, 1554 (1993); 
H.\ Okawa,\newline
\hspace*{0.49 cm} N.\ Matsumoto, H.\ Tamaki and M.\ Obba, 
Mol.\ Cryst.\ Liq.\ Cryst.\ \newline
\hspace*{0.49 cm} {\bf 233}, 257 (1993);
M.\ Turnbull, C.~P.\ Landee, T.~C.\ Soesbe and R.~D.\ Willet,
\newline
\hspace*{0.49 cm} Mol.\ Cryst.\ Liq.\ Crys.\ {\bf 233}, 269 (1993).
\vskip 3 truept
\noindent {\bf 2.} L.~N\'eel, Ann.\ Phys., Paris {\bf 3}, 137 (1948).
\vskip 3 truept
\noindent {\bf 3.} M.~Mansuripur, J.\ Appl.\ Phys.\ {\bf 61}, 1580 (1987); 
\newline 
\hspace*{0.49 cm} F.~Tanaka, S.~Tanaka, N.~Imanura, 
Jpn.\ J.\ Appl.\ Phys.\ {\bf 26}, 231 (1987).
\vskip 3 truept
\noindent {\bf 4.} A.F.\ Siqueira and I.P.\ Fittipaldi, 
J.\ Magn.\ Magn.\ Mater.\ {\bf 54}, 678 (1986); \newline
\hspace*{0.49 cm} A.~Bobak and M.~Jascur, Phys.\ Rev.\ B {\bf 51}, 11533 (1995).
\vskip 3 truept
\noindent {\bf 5.} G.M.\ Buend{\y}a and M.A.\ Novotny, 
J.\ Phys.: Condens.\ Matter {\bf 9}, 5951 (1997).
\vskip 3 truept
\noindent {\bf 6.} G.M.\ Buend{\y}a, M.A.\ Novotny and J.~Zhang, in 
\underline{Computer Simulations in Condensed} \newline 
\hspace*{0.49 cm} \underline{Matter Physics VII}, 
ed.\ D.P.\ Landau, K.K.\ Mon and H.-B.\ Sch\"uttler (Springer, Berlin, \newline
\hspace*{0.49 cm} 1994), p.~223; G.M.\ Buend{\y}a and J.A.\ Liendo,
J.\ Phys.: Condens.\ Matter {\bf 9}, 5439 (1997).
\vskip 3 truept
\noindent {\bf 7.} K.~Binder in 
\underline{Monte Carlo Methods in Statistical Physics}, ed.\ K.~Binder \newline 
\hspace*{0.49 cm} (Springer, Berlin, 1979).

\end{document}